\documentclass[floatfix
 reprint,
]{revtex4-2}

\usepackage{graphicx}
\usepackage{dcolumn}
\usepackage{bm}
\usepackage{amsmath,amsfonts}
\usepackage{xcolor}

\begin{document}

\preprint{APS/123-QED}

\title{Extensive Composable Entropy for the Analysis of   Cosmological Data} 

\author{Constantino Tsallis}
\affiliation{Centro Brasileiro de Pesquisas F\'{\i}sicas and National Institute of Science and Technology of Complex Systems, Rua Xavier Sigaud 150, 22290-180, Rio de Janeiro RJ, Brazil,}
\affiliation{Santa Fe Institute, 1399 Hyde Park Road, Santa Fe, 87501 NM, USA,}
\affiliation{Complexity Science Hub Vienna,  Josefst\"adter Strasse 39, 1080 Vienna, Austria}

\author{Henrik Jeldtoft Jensen}
 
\affiliation{Centre for Complexity Science and Department of Mathematics, Imperial College
London, South Kensington Campus, London, SW7 2AZ, United Kingdom,}
\affiliation{School of Computer Science, Tokyo Institute of Technology, 4259,
Nagatsuta-cho, Yokohama 226-8502, Japan,}
\affiliation{Complexity Science Hub Vienna,  Josefst\"adter Strasse 39, 1080 Vienna, Austria}

\date{\today}
            
\begin{abstract}
Along recent decades, an intensive worldwide research activity is focusing both black holes and cosmos (e.g. the dark-energy phenomenon) on the basis of entropic approaches. The Boltzmann-Gibbs-based Bekenstein-Hawking entropy $S_{BH}\propto A/l_P^2$ ($A \equiv$ area; $l_P \equiv$ Planck length) systematically plays a crucial theoretical 
role although it has a serious drawback, namely that it violates the thermodynamic extensivity of spatially-three-dimensional systems. Still, its intriguing area dependence points out the relevance
of considering the form $W(N)\sim \mu^{N^\gamma}\;\;(\mu >1;\gamma >0)$, $W$ and $N$ respectively being the total number of microscopic possibilities and the number of components; $\gamma=1$ corresponds to standard Boltzmann-Gibbs (BG) statistical mechanics. For this $W(N)$ asymptotic behavior, we introduce here,
on a group-theory basis, the entropic functional 
$S_{\alpha,\gamma}=k \Bigl[ \frac{\ln \Sigma_{i=1}^W p_i^\alpha}{1-\alpha} \Bigr]^{\frac{1}{\gamma}} \;(\alpha \in 
\mathbb{R};\,S_{1,1}=S_{BG}\equiv-k\sum_{i=1}^W p_i \ln p_i)$. 
This functional simultaneously is {\it extensive} (as required by thermodynamics) and {\it composable} (as required for logic consistency), 
$\forall (\alpha,\gamma)$. We further show that $(\alpha,\gamma)=(1,2/3)$ satisfactorily agrees with cosmological data measuring neutrinos, Big Bang nucleosynthesis and the relic abundance of cold dark matter particles, as well as dynamical and geometrical cosmological data sets.
\end{abstract}

\maketitle

Recent decades have witnessed intense activity along entropic lines focusing on elusive cosmological issues (e.g., dark energy and dark matter) and black holes. They connect thermodynamics, the so-called area law, and diverse holographic models. An important milestone is constituted by the Bekenstein-Hawking entropy $S_{BH}$ and its thermodynamical consequences \cite{Bekenstein1973,Bekenstein1974,Hawking1974,Hawking1976}. It has been established on solid grounds that $S_{BH}$, based on
the Boltzmann-Gibbs (BG) entropy $S_{BG}\equiv -k\sum_{i=1}^W p_i \ln p_i$ ($S_{BG}=k \ln W$ for the particular instance of equal probabilities), is proportional to the horizon area instead of the associated volume of the black hole. The physical reason is convincingly related to the holographic content of information of such heavily gravitational systems.

However, $S_{BH}$ is not extensive (i.e., proportional to the number of constituents or equivalent geometrical quantities such as the volume), which appears to be at odds with standard thermodynamics. This intriguing aspect of $S_{BH}$ has been lengthily emphasized in many articles, e.g., \cite{tHooft1985,tHooft1990,Susskind1993,Maddox1993,Srednicki1993,StromingerVafa1996,MaldacenaStrominger1998,DasandShankaranarayanan2006,BrusteinEinhornYarom2006,BorstenDahanayakeDuffEbrahimRubens2009,Padmanabhan2009,Casini2009,BorstenDahanayakeDuffMarraniRubens2010,Corda2011,KolekarPadmanabhan2011,Saida2011}. 

In principle one may call any functional from spaces of probability distributions to the real numbers for an entropy, but it seems mathematically and logically appealing to build entropic functionals on an axiomatic foundation. It is well known that the BG functional form is the only one possible if one assumes the four Shannon-Khinchin (SK) axioms, see e.g. \cite{Tempesta_Annals_2016}. The BG entropy is only extensive when the number of available micro-configurations $W(N)$ of the system's $N$ constituents is exponential in $N$. In \cite{TsallisCirto2013} it was pointed out that the so-called area-law is, for $d>1$ dimensions, consistent with the following asymptotic behavior 
\begin{equation}
W(N)\sim A e^{N^\gamma} \;\;\;(N\to \infty; \, A>0; \gamma >0)\,.
\label{stretched}
\end{equation}
This means that we need an entropy which is extensive for this stretched exponential growth rate of the number of available configurations and which also satisfies a set of basic axioms. The group entropies introduced by Tempesta in \cite{Tempesta_PRE_2011,TempestaJensen2020} offers such a framework. The first three SK axioms together with a fourth axiom replacing the additive SK axiom by a more general composability axiom \cite{Tsallis2009,Tsallis2023} are satisfied by group entropies generated from formal group generators. 

Before proceeding, let us remind at this point the precise definition of entropic composability, which consists in considering two probabilistically independent systems $A$ and $B$ as 
one single
system $A \times B$ with a set of possible states given by the Cartesian combination of the states for $A$ and for $B$. We then require
\begin{equation}
\frac{S(A \times B)}{k}= F\Bigl(\frac{S(A)}{k},\frac{S(B)}{k}\Bigr) \,,   
\end{equation}
where $F(x,y;\{\eta\})$ is a smooth function of $(x,y)$ which depends on a (typically small) set of universal indices $\{\eta\}$ defined in such a way that $F(x,y; \{ 0\})=x+y$ ({\it additivity}), and which satisfies $F(x,0;\{\eta\})=x$ ({\it null-composability}), $F(x,y;\{\eta\})=F(y,x;\{\eta\})$ ({\it symmetry}), $F(x,F(y,z;\{\eta\}); \{ \eta \})=   F(F(x,y;\{\eta\}),z;\{\eta\})$ 
({\it associativity}). 
For thermodynamical systems, this associativity appears to be consistent with the $0th$ Principle of Thermodynamics. A general proof is not yet available, but the entropic functional $S_q \equiv k \frac{1-\sum_{i=1}^W p_i^q}{q-1}\;(q \in \mathbb{R}; S_1=S_{BG})$ (introduced in \cite{Tsallis1988} in order to generalize Boltzmann-Gibbs statistical mechanics) is composable with $F(x,y)=x+y+(1-q)\,x\,y$, and it has been explicitly shown to satisfy the $0th$ Principle for many-body overdamped systems with two-body repulsive interactions \cite{AndradeSilvaMoreiraNobreCurado2010,NobreCuradoSouzaAndrade2015,VieiraCarmonaAndradeMoreira2016,SouzaAndradeNobreCurado2018}.
$S_q$ is in fact the unique entropic functional which simultaneously satisfies the trace-form and composability properties \cite{EncisoTempesta2017}. This entropic functional for $q \ne 1$ is definitively appropriate for systems belonging to the family $W(N) \sim B N^\rho \;\;(N\to\infty;\,B>0;\,\rho >0)$, but it turns out to be  inadequate for generically handling the systems belonging to the family $(\ref{stretched})$, which constitutes our present focus. 

If one insists on composability for all probability distributions (and not just for the uniform distribution) one is forced, for family $(\ref{stretched})$,   to abandon the trace form, as for example satisfied by the $S_{BG}$ entropy, and have instead an infinity of non-trace forms generalising in highly non-trivial ways the R{\'e}nyi entropy
$S_\alpha^R[p] \equiv k\frac{\ln \sum_{i=1}^W p_i^\alpha}{1-\alpha}\;(S_1^R=S_{BG})$.
The four axioms  introduced to make a secure logical foundation for the group entropies do not uniquely determine the functional form of the entropy, i.e. the situation is different from the SK case. However, if one adds to the four axioms the additional requirement that the entropy must be extensive for the relevant functional form of $W(N)$, one is able to express the group generator for the corresponding group entropy in terms of the inverse of $W(N)$, see e.g. \cite{JensenTempesta2024}. We describe now how to derive a group  entropy which is extensive for the case  $W(N)\sim A e^{N^\gamma}$.

The non-trace form group entropies are given in terms of the group generator $t\mapsto G(t)$ by the expression \cite{JensenTempesta2024}
\begin{equation}
    S[p]=G\Bigl(\frac{\ln \sum_{i=1}^W p_i^\alpha}{1-\alpha} \Bigr)\;\; {\rm with}\;\; \alpha >0\;\; {\rm and}\;\; \alpha \neq 1.
\end{equation}
The group generator is determined by the inverse of $W(N)$ and given by \cite{JensenTempesta2024}
\begin{equation}
    G(t) = \lambda(1-\alpha)\{ W^{-1}[\exp(\frac{t}{1-\alpha})]-W^{-1}(1)\}.
\end{equation}
Here $\lambda \in \mathbb{R}_{+}$ is a free parameter.

The inverse of the function in Eq. \ref{stretched} is given by 
\begin{equation}
    W^{-1}(t)=[\ln t]^{\frac{1}{\gamma}},
\end{equation}
and we arrive at the following expression for the corresponding group entropy 
\begin{equation}
    S_{\alpha,\gamma}[p]=\lambda\Bigl[\frac{\ln \sum_{i=1}^W p_i^\alpha}{1-\alpha}\Bigr]^{\frac{1}{\gamma}}.
    \label{alphagammaentropy}
\end{equation}
The factor $\lambda$ is an as yet unspecified constant, like the Boltzmann constant $k$, and can be taken $\lambda=k$ or, for simplicity, $\lambda=1$. The expression within the square brackets is the R{\'e}nyi entropy, which, as indicated above, in the limit $\alpha \rightarrow 1$ reduces to the BG entropy. So, if guided by simplicity, we assume this limit, we obtain 
\begin{equation}
    S_{1,\gamma}[p] = (S_{BG}[p])^\frac{1}{\gamma}.
\label{newentropy}    
\end{equation}
We emphasise that this entropic form satisfies the first three SK axioms, the composability axiom of Tempesta and is extensive 
in the case of the area law. The fact that the entropy is extensive suggests this entropy could be the relevant thermodynamic entropy. It is not immediately clear if the temperature defined as $\partial E/\partial S_{1,\gamma}$ together with a maximum entropy approach under the constraint of the average energy $E$ will satisfy the zeroth law of thermodynamics. Also, more work is needed to fully understand to what extent the usual Legendre structure of thermodynamics is satisfied by a statistical mechanics based on maximising the entropy $S_{1,\gamma}[p]$ to obtain the probability weights.

We need now to determine the exponent $\gamma$.

If the physical system satisfies the so-called {\it area law}, i.e., if $S_{BG}\propto L^{d-1}$ ($d>1$) where $L$ is the linear size of a $d$-dimensional system, then $\gamma=(d-1)/d$, hence 
\begin{equation}
\delta= \frac{1}{\gamma} = \frac{d}{d-1} \,.
\end{equation}

We recall that, for a $d=3$ system such as a physical black hole, or possibly the universe as a whole, the so-called $\delta$-entropic functional 
\begin{equation}
S_\delta [p]\equiv k \sum_{i=1}^W p_i \Bigl[\ln \frac{1}{p_i} \Bigr]^\delta
\label{deltaentropy}
\end{equation}
was conjectured based on clear physical arguments in \cite{Tsallis2009,Tsallis2023} with $\delta=3/2$, making the $\delta$-entropy to neatly differ from the Bekenstein-Hawking entropy $S_{BH}$   which corresponds to $\delta=1$. Although the trace-form entropy $S_\delta [p]$ is extensive for the relevant $W(N)$ dependence in Eq. (\ref{stretched}), it is not fully composable. It is unclear, at the present stage, whether composability is necessary for the $0$-th principle to be satisfied for (conservative and/or dissipative) many-body systems. On the other hand, in this sense, the non-trace-form $S_{1,\gamma}[p]$ constitutes a strong potential candidate for black-hole thermodynamics as it is both composable and extensive for $W(N)$ in Eq. (\ref{stretched}).

In terms of the Bekenstein-Hawking entropy, we have that
\begin{equation}
S_{1,\gamma}=k \Bigl[ \frac{S_{BH}}{k} \Bigr]^{\frac{1}{\gamma}} \,.
\end{equation}
is thermodynamically extensive for $\gamma=1/\delta=2/3$.

\begin{figure}
\includegraphics[width=0.47\textwidth]{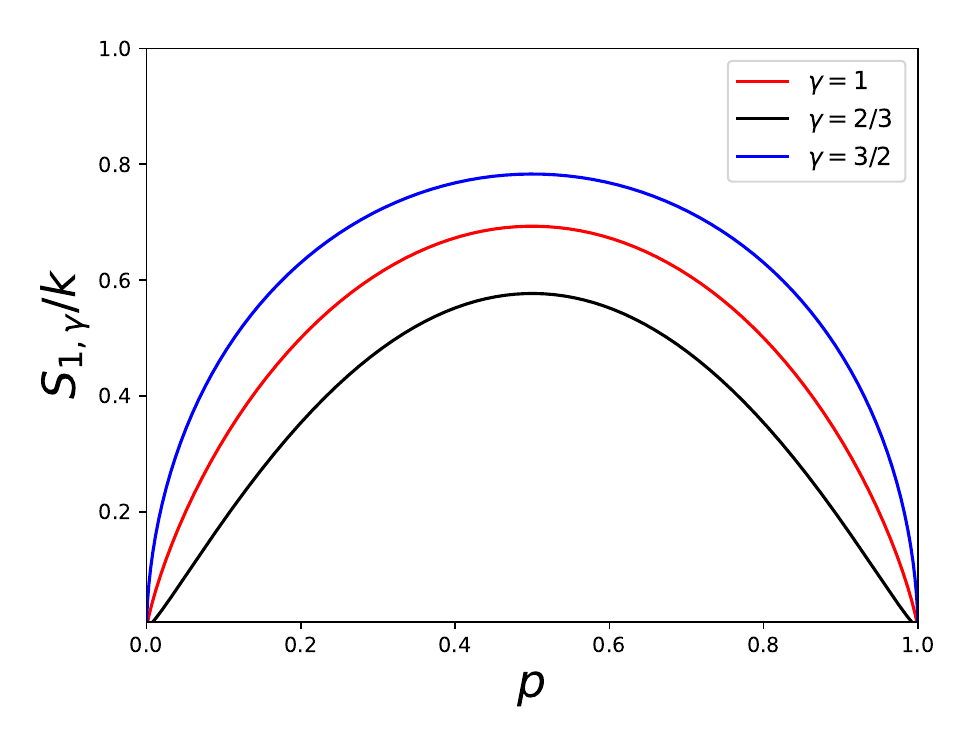}
\caption{\label{Sqgamma}Entropic functional $S_{1,\gamma}$, as given by Eq. (\ref{newentropy}), for the binary particular case ($W=2$), with typical values of $\gamma$. Concavity is lost for $\gamma$ low enough.}
\end{figure}
\begin{figure}
\includegraphics[width=0.47\textwidth]{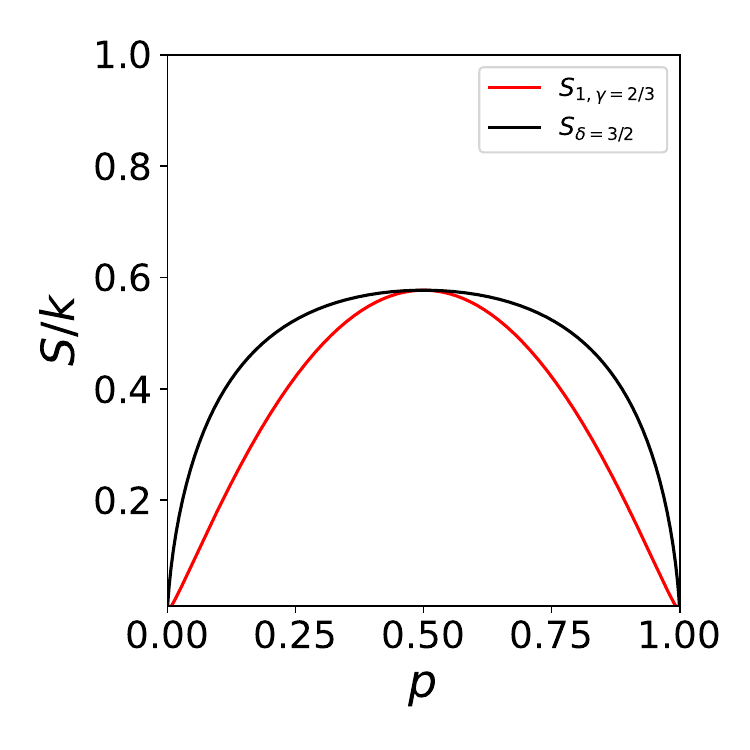}
\caption{\label{binary}The entropic functionals  $S_{1,\gamma}$ (Eq. (\ref{alphagammaentropy}); extensive,  composable, and non-trace-form) and $S_\delta$ (Eq. (\ref{deltaentropy}; extensive, trace-form, and non-composable); extensive, non) for the binary particular case ($W=2$) that are extensive for the family (\ref{stretched}); notice that their values coincide for equal probabilities and are given by $(\ln 2)^{3/2}$. Besides these, there are other entropic functionals which are related with stretched-exponential functions, such as \cite{AnteneodoPlastino1999,HanelThurner2011a,HanelThurner2011b}.}
\end{figure}

The Barrow entropy $S_B$ \cite{Barrow2020} deserves a comment at this point. Let us start by clarifying that it is not an entropic functional since no particular dependence on probabilities is proposed. It basically consists in considering the possibility of the surface of a black hole being not smooth but a fractal instead. It can be defined as follows:
\begin{equation}
S_B \propto [S_{BH}]^{1 + \frac{\Delta}{2}} \,\, (0 \le \Delta \le 1) \, . 
\end{equation}

Since both $S_\delta$ and $S_{1,\gamma}$ yield, for the microcanonical ensemble (i.e., for the equal probabilities case) (see Figs. \ref{Sqgamma} and \ref{binary}),
\begin{equation}
S_\delta = S_{1,\gamma} \propto [S_{BH}]^\delta \,,
\end{equation}
we can generically identify
\begin{equation}
\delta=1/\gamma=1+ \frac{\Delta}{2} \,.
\end{equation}

Most important, let us now compare the present extensive and hence potentially thermodynamically consistent theoretical prediction $\delta=1/\gamma = 3/2$ with available observational results \cite{JizbaLambiase2022,SalehiPouraliAbedini2023,JizbaLambiase2023,DenkiewiczSalzanoDabrowski2023}, having in mind that the Bekenstein-Hawking proposal corresponds to $\delta=1/\gamma=1$ (i.e. $\Delta=0$).\\ 
(i) In the context of dark matter, the high-energy neutrino data collected at the IceCube Neutrino Observatory (South Pole) yield \cite{JizbaLambiase2022}  $\delta =1.565$, amazingly close to $\delta=3/2$, thus excluding $\delta=1$. \\
(ii) The neutrino data collected at the Planck Observatory (ESA) have been discussed in \cite{SalehiPouraliAbedini2023} through two different theoretical approaches: one of them yields $\delta=1=1+ \frac{\Delta}{2}=1.87$, whereas the other one yields $1.26$. By conjecturally assuming that truth is somewhere in the middle (e.g., if these two theoretical approaches yield kind of upper and lower bounds of the correct value), we may verify that $\delta \simeq \frac{1.87+1.26}{2}=1.565$, intriguingly enough, the same value just mentioned above! \\
(iii) The Bing Bang nucleosynthesis is focused on in \cite{JizbaLambiase2023} and the relic abundance analysis of cold dark matter particles yields $\delta \simeq 1.499$, once again amazingly close to $\delta=3/2$. \\ 
(iv) Finally, through the analysis of dynamical and geometrical cosmological data sets, it is obtained  in \cite{DenkiewiczSalzanoDabrowski2023} that 
$\delta=1+ \frac{\Delta}{2}>1.43$. 

Summarizing, observations are neatly favorable to $\delta=3/2$, as first suggested in \cite{TsallisCirto2013}, as being, for such heavily gravitational systems, the thermodynamically correct one,  and they exclude the Bekenstein-Hawking entropy, constructed upon the Boltzmann-Gibbs one. Further observational data
concerning phenomena neatly unrelated to equal probabilities are needed to discriminate between $S_{1,\gamma}[p]$ and $S_\delta[p]$ as being the physically admissible entropic functional for such gravitational systems. Both are extensive for $\delta=1/\gamma=3/2$, but whereas $S_\delta[p]$ is trace-form but not composable, $S_{1,\gamma}[p]$ is composable but not trace-form.

The above issues echo the following lines in the celebrated 1902 book by Gibbs \cite{Gibbs1902}: 
{\it In treating of the canonical distribution, we shall always suppose the multiple integral in equation (92)} [the partition function, as we call it nowadays] {\it to have a finite value, as otherwise the coefficient of probability vanishes, and the law of distribution becomes illusory. This will exclude certain cases, but not such apparently, as will affect the value of our results with respect to their bearing on thermodynamics. It will exclude, for instance, cases in which the system or parts of it can be distributed in unlimited space} [...]. {\it It also excludes many cases in which the energy can decrease without limit, as when the system contains material points which attract one another inversely as the squares of their distances.} [...]. {\it For the purposes of a general discussion, it is sufficient to call attention to the assumption implicitly involved in the formula (92).}

We acknowledge useful remarks from H.S. Lima. One of us (CT) also acknowledges partial financial support by CNPq and Faperj (Brazilian Agencies).

\end{document}